\documentclass[preprint,12pt]{elsarticle}

\usepackage{amssymb}
\usepackage{lipsum}
\usepackage{amsmath}
\usepackage{needspace}
\usepackage{booktabs} 
\usepackage{array} 
\usepackage{cuted}

\journal{Physics Letters B}

\begin{document}

\begin{frontmatter}



\title{Rotational mass of anisotropic neutron stars within Rastall gravity}

\def\Journal#1#2#3#4{{ #1} {\bf #2}, #3 (#4) }
\def\RPP{{Rep. Prog. Phys}}
\def\PRC{{Phys. Rev. C}}
\def\PRD{{Phys. Rev. D}}
\def\ZPA{{Z. Phys. A}}
\def\NPA{{Nucl. Phys. A}} 
\def\JPG{{J. Phys. G }}
\def\PRL{{Phys. Rev. Lett}}
\def\PR{{Phys. Rep.}}
\def\PREV{{Phys. Rev.}}
\def\PRX{{Phys. Rev. X}}
\def\PLB{{Phys. Lett. B}}
\def\AP{{Ann. Phys (N.Y.)}}
\def\EPJA{{Eur. Phys. J. A}}
\def\NP{{Nucl. Phys}}  
\def\RMP{{Rev. Mod. Phys}}
\def\IJMPE{{Int. J. Mod. Phys. E}}
\def\AJ{{Astrophys. J}}
\def\AJL{{Astrophys. J. Lett}}
\def\AA{{Astron. Astrophys}}
\def\ARAA{{Annu. Rev. Astron. Astrophys}}
\def\MPLA{{Mod. Phys. Lett. A}}
\def\ARNPS{{Annu. Rev. Nuc. Part. Sci}}
\def\LRR{{Living. Rev. Relativity}}
\def\CARAA{{Class. Ann. Rev. Astron. Astrophys.}}
\def\EPJC{{Eur. Phys. J. C}}
\def\cqg{{Class. Quantum Grav.}}
\def\mon{{Mon. Not. R. Astron. Soc.}}
\def\AJSS{{Astrophys. J. Suppl. Ser.}}

 \author[label1]{M. Lawrence Pattersons}
 \ead{m.pattersons@proton.me}
 \author[label1,label2]{Freddy P. Zen}
 \ead{fpzen@fi.itb.ac.id}
 \author[label3,label2]{Hadyan L. Prihadi}
 \ead{hady001@brin.go.id}
 \author[label4]{Muhammad F. A. R. Sakti}
 \ead{fitrahalfian@gmail.com}
 \affiliation[label1]{organization={Theoretical High Energy Physics Group, Department of Physics, Institut Teknologi Bandung},
             addressline={Jl. Ganesha 10},
             city={Bandung},
             postcode={40132},
             country={Indonesia}}
\affiliation[label2]{organization={Indonesia Center for Theoretical and Mathematical Physics (ICTMP), Institut Teknologi Bandung},
             addressline={Jl. Ganesha 10},
             city={Bandung},
             postcode={40132},
             country={Indonesia}}
 \affiliation[label3]{organization={Research Center for Quantum Physics, National Research and Innovation Agency (BRIN)},
             city={South Tangerang},
             postcode={15314},
             country={Indonesia}}
 \affiliation[label4]{organization={High Energy Physics Theory Group, Department of Physics, Faculty of Science, Chulalangkorn University},
             city={Bangkok},
             postcode={10330},
             country={Thailand}}


\begin{abstract}
Due to rotation, the mass correction of neutron stars arises and causes the rotational mass to be larger than the static mass. In this work, we extend the formulation of the rotational mass of anisotropic neutron stars within Rastall gravity. We apply numerical simulation on the formulation we obtained. We refer to mass of J0740+6620, GW170817, and GW190814 as the mass constraints of the neutron stars. For the free parameters, we use three values of Rastall's parameter, i.e. $\lambda=0.00019$, $\lambda=0.00038$, $\lambda=0.00071$; and three values of anisotropic strength, i.e. $\zeta=-1.15$, $\zeta=-1.50$, and $\zeta=-2.00$. We have found that both $\lambda$ and $\zeta$ impact on the increment of the NS's rotational mass within the compact regimes, and also impact on the decrease of the NS's rotational mass within the loose regimes. All mass constraints are satisfied by the NS with $\zeta=-2.00$. In term of the moment of inertia $I$ of the neutron stars, all numerical results match with the constraint range which is based on radio observations of heavy pulsars; while in term of the angular velocity of the stars relative to the distant observers $\Omega$, the mass correction $\delta M$ significantly increases when $\Omega$ increases. 
\end{abstract}




\end{frontmatter}




\section{Introduction}
\label{introduction}

It is widely known that the mass of static neutron stars (NSs) can be obtained by solving Tolman-Oppenheimer-Volkoff (TOV) equation \cite{Opp}. When we proceed to study the rotating NSs, we have to calculate the mass correction which arises in the formulation. Rotational mass of rotating relativistic stars within general relativity (GR) was proposed by Hartle \& Thorne \cite{Har1,Har2}. In Hartle-Thorne (HT) formalism, rotation is handled as a perturbation for static configuration of the stars. As a direct consequence, the metric and the energy-momentum tensor (EMT) in HT formalism are expanded in multipole expansion, which is truncated at quadrupole order. HT formalism has been largely used by many authors (see Refs. \cite{Wen,Weber,weber2,Ordaz,Urb,Gup,Pin,Bez,Moreno,Lopes}), mostly concerning NSs \cite{Wen,Weber,weber2,Ordaz,Ber,Sil,Pat}.

{Berti et al. \cite{Ber} conducted a comparative analysis of the HT approximation, the Cook–Shapiro–Teukolsky (CST) solutions \cite{Cook}, and the Manko–Mi-elke–Sanabria-Gómez (MMSG) formalism \cite{Mank}. Note that the CST framework accommodates both slowly and rapidly rotating compact stars. In the comparison between the HT approximation and the CST models, configurations were selected such that the gravitational mass and angular momentum were matched. Their results demonstrate that the HT formalism provides remarkably accurate predictions for the corotating and counterrotating radii, $R_\pm$, of the innermost stable circular orbit, with deviations less than 1 \% even for the most rapidly rotating millisecond pulsars. When the HT approximation is compared to the MMSG formalism—specifically in the exterior vacuum spacetime—it also demonstrates better overall performance and higher accuracy.}

Pattersons \& Sulaksono \cite{Pat} have extended HT formalism for NSs whose anisotropy, i.e. unequal value between radial and tangential pressures inside NS matter. The anisotropy could be generated by many factors due to matter properties which are linking to each other such as boson condensations, the existence of solid core, different kinds of phase transition, as well as the presence of strong magnetic and electric field \cite{Pat,Sul1,Her}; or due to the impact of modified gravity, see Refs. \cite{Danarianto2019,Wojnar2016,Ovalle2017,Afonso2018a,Afonso2018b}. Recently, Beltracchi \& Posada \cite{Bel}, reformulate HT formalism for slowly rotating anisotropic relativistic stars. They also revealed some mistypes and errors in Ref. \cite{Pat}. Even so, this magnificent reformulation purely based on GR and does not involve any modified gravity theory.\\
\indent On the other hand, if GR is correct, there are some problems left to be explained, e.g. dark matter and dark energy \cite{Clifton2012}. An alternative way out of these problems is considering the modified theories of gravity. One of the examples of modified gravity is the theory constructed by Rastall \cite{Ras}. It is claimed as a generalization of GR. In his paper, he proposed an idea that the covariant divergence of EMT does not vanish, but it is proportional to the divergence of the Ricci Scalar, i.e. $\nabla_\mu T^{\mu\nu}\varpropto\nabla^\nu \mathcal{R}$. This condition implies the violation of EMT conservation, which can be phenomenologically caused by quantum effects in classical context \cite{Oli,Sak}.\\
\indent Visser \cite{Vis} claimed that Rastall gravity is apparently equivalent to GR. He stated that the Rastall EMT, which is generated in the Rastall field equation, is merely an artificially isolated part of the physical conserved EMT. On the other side, Darabi et al. \cite{Dar} refuted Visser's view. They emphasized that Rastall did not introduce some new EMTs in his original paper. In fact, Rastall proposed a new idea on the relation between EMT and geometry which is supported by an unknown mutual interaction between them. They also stressed that the violation of energy conservation has been invoked in frameworks which are different from Rastall gravity approach, e.g. in an attempt to achieve dark energy (see Ref. \cite{Dar} to see their full refutation on Visser's view). Beside that, many studies \cite{Oli,Das,Sak2,Men,Tan,Pre,Nas2,Nas,Tan2,Sal,Xi} have shown that Rastall gravity is actually different from GR. Moreover, Rastall gravity has been involved in numerous studies regarding various objects, such as black holes (BHs) \cite{Sak,Sak2,Nas,Pri,Spa,Hey,Zou,Kumar,Heydar}, wormholes \cite{Bha,Malik,Hal,Mor,Hey2,Bro}, strange quark stars \cite{Sal,Aya}, and gravastar \cite{Maj}. Also, Rastall gravity has been applied in the study of accelerated expanding universe \cite{Cap}.\\
\indent Since the essence of Rastall gravity is associated to high curvature environments, NSs seem to be promising laboratories to test it \cite{Oli}. In the context of NSs in Rastall gravity, several studies have been done. Oliveira et al. \cite{Oli} obtained the formulation of spherically symmetric isotropic NSs, and they also obtained its numerical solutions; while Xi et al. \cite{Xi} worked on the same case, but they used a different equation of state (EOS). Da Silva et al. \cite{Das} constructed the formulation of rotating isotropic NSs. Meng \& Liu \cite{Men} derived the tidal Love numbers of NSs. Majeed et al. \cite{Maj2} modeled static anisotropic compact stars. However, NSs' gravity has not been fully understood yet \cite{Pat}. This fact and the studies on NSs in Rastall gravity that have been mentioned above give us opportunity to explore the application of Rastall gravity on NSs.\\
\indent In 2019, the LIGO/Virgo collaboration observed GW190814 of the merger of a BH with mass of (22.2$-$24.3) $M_\odot$ and a secondary compact object with mass of (2.50$-$2.67) $M_\odot$ \cite{Abbott2020}. The secondary object could be the lightest BH or the heaviest NS that has been observed. Up to now, this problem is not really clear. Some authors \cite{Cap2,Tew,Cle,Lyu} tend to assume that it was a BH, while some others \cite{Hua,Dex,Bis} assume that it was a NS. The work conducted in Ref. \cite{Pat} could not achieve the mass range of the secondary object of GW190814, even though they have impose anisotropy on NSs within HT procedure.\\
\indent Motivated by the previous studies, in this work, we re-derive the formalisms of anisotropic TOV equation and rotational mass of NSs formulated in Ref. \cite{Pat} and \cite{Bel} within Rastall gravity. We use the anisotropic model proposed by Horvat et al. \cite{Hor}. The model assumes that the difference between radial pressure $p$ and tangential pressure $q$ is proportional to the radial pressure $p$ and the mass at some distance $r$ from the center of the star $m(r)$, and inversely proportional to the distance from the center. For the EOS, we use the EOS for NS matter with nuclei in the crust and hyperons in the core, proposed by Miyatsu et al. \cite{Miy}. We use three mass constraints in this study, i.e. J0740+6620 \cite{Fonseca2021} with mass of $2.08^{+0.07}_{-0.07} M_\odot$, GW170817 \cite{Abbottgw170817} whose heaviest object with mass in the range of 1.36$-$2.26 $M_\odot$, and GW190814 \cite{Abbott2020} whose secondary object with mass in the range of 2.50$-$2.67 $M_\odot$. The highest mass constraint is achieved by GW190814. The main distinction between our work and previous works which include rotating NS (such as \cite{Wen,Pat,Rah,Prasetyo}) is that our result depends on the Rastall's parameter. We find that the Rastall's parameter influences mass-radius relation of the NS, as well as the anisotropic strength highly influences the rotational mass increment. Such a mass increment could be achieved in Ref. \cite{Das}, but the EOS we use is more realistic than the one they used. Our results strengthen the argument that the secondary compact object of GW190814 is a NS.\\
\indent This paper is organized as follows. In section 2, we explain the construction of formalisms we use. In section 3, we discuss the numerical results. Lastly, section 4 contains summary and conclusions of this work.

\section{Formalisms}
\label{sec:1}
In this section, we divide the formalism {four} parts. In sub-section \ref{sec:2}, we briefly review the Rastall gravity formalism. {The nuclear model for the EOS is shortly discussed in sub-section~\ref{EOS}}. The anisotropic NSs' rotational mass formalism within Hartle-Thorne-Rastall framework is given in sub-section~\ref{sec:3}. The profile of anisotropic pressure is described in sub-section~\ref{sec:4}. It is worth noting that we use geometrized units $G=c=1$ in this paper.

\subsection{Rastall gravity}
\label{sec:2}
It has been widely known that in GR, the covariant divergence of EMT vanishes, i.e.
\begin{equation}
\nabla_\mu T^{\mu\nu}=0,
\end{equation}
while in Rastall gravity, the covariant divergence of EMT can be expressed as \cite{Sak} 
\begin{equation}
\nabla_\mu T^{\mu\nu}=\lambda \nabla^{\nu}\mathcal{R}.
\label{covdivRastall}
\end{equation}
here $\lambda$ is a parameter which determines the deviation from general relativity, and $\mathcal{R}$ is Ricci scalar. Eq. (\ref{covdivRastall}) results in the emergence of a new field equation that is different from Einstein field equation. The new field equation writes
\begin{equation}
G^\mu_\nu+\kappa\lambda \mathcal{R} \delta^\mu_\nu = \kappa T^\mu_\nu.
\label{fieldeq}
\end{equation}
From the Rastall field equation shown by Eq. (\ref{fieldeq}), for 4-dimensional spacetime, we can obtain a relation
\begin{equation}
\mathcal{R}=\frac{\kappa T}{4\kappa\lambda - 1},
\label{RicciRastall}
\end{equation}
where $T$ is the energy-momentum scalar.

Now Eq. (\ref{fieldeq}) can be simplified as the following
\begin{equation}
G^\mu_\nu=\kappa \tilde{T}^\mu_\nu,
\label{fieldeqRastall}
\end{equation}
where
\begin{equation}
\tilde{T}^\mu_\nu=T^\mu_\nu-\delta^\mu_\nu\left(\frac{\kappa\lambda}{4\kappa\lambda-1}T\right).
\label{Teffective}
\end{equation}

It is well known that in GR, $\kappa=8\pi$. Yet it is worthy to note that the value is attained by applying GR to the weak-field limit. By applying Rastall gravity in the weak-field limit, da Silva et al. \cite{Das} aprroximate $\kappa$ as 
\begin{equation}
\kappa=\frac{8\pi}{2\Lambda+1},
\label{kappa}
\end{equation}
where $\Lambda$ satisfies
\begin{equation}
    \lambda=\frac{\Lambda}{\kappa(4\Lambda-1)}
    \label{capitallambda}
\end{equation}

{Ref. \cite{Moradpour2016} provides an additional important informations regarding allowed values of the Rastall parameter. The $\kappa$ has to satisfy}
\begin{equation}
    {\kappa=\frac{4\kappa\lambda-1}{6\kappa\lambda-1}8\pi,}\label{kappamoradpour}
\end{equation}
{and the $\lambda$ has to satisfy}
\begin{equation}
    {\lambda=\frac{\kappa\lambda(6\kappa\lambda-1)}{(4\kappa\lambda-1)8\pi}.}\label{lambdamoradpour}
\end{equation}

From Eq. (\ref{RicciRastall}), we could take an insight that $\kappa\lambda=1/4$ mathematically gives us a singularity, so this value is forbidden. From Eq. (\ref{kappa}) and (\ref{capitallambda}), we can see that $\Lambda=-1/2$ and $\Lambda=1/4$ are also forbidden. {Moreover, Eq. (\ref{kappamoradpour}) conveys the same information as Eq. (\ref{RicciRastall}), while Eq. (\ref{lambdamoradpour}) reveals that the condition $\kappa\lambda=1/6$ is also not allowed.} We can verify that, as we expect, Rastall gravity recovers to GR when we set $\lambda=0$.\\
\subsection{{Nuclear model of the EOS}} \label{EOS}
{The Lagrangian of the nuclear model used in this work is given by $\mathcal{L}=\mathcal{L}_B+\mathcal{L}_M+\mathcal{L}_{\text{int}}$ \cite{Miy}.}

{Here $\mathcal{L}_B$ denotes the baryon term which reads}
\begin{eqnarray}
    {\mathcal{L}_B=\sum_B \bar{\psi}_B(i\gamma_\mu\partial^\mu-M_B)\psi_B,} \label{lagragianb}
\end{eqnarray}
{where $\psi_B$ is the baryon field and $M_B$ is the baryon mass in vacuum. Note that the sum $B$ runs over the octet baryons $p$, $n$, $\Lambda$, $\Sigma^{+0-}$, and $\Xi^{0-}$. Here $M_B$ denotes baryon mass.}

{The Lagrangian of meson term $\mathcal{L}_M$ reads}
\begin{eqnarray}
{\mathcal{L}_M} &=& 
{\frac{1}{2} \left( \partial_\mu \sigma \, \partial^\mu \sigma - m_\sigma^2 \sigma^2 \right)} 
+ {\frac{1}{2} m_\omega^2 \omega_\mu \omega^\mu} 
- {\frac{1}{4} W_{\mu\nu} W^{\mu\nu}} \nonumber \\
&& 
{+ \frac{1}{2} m_\rho^2 \, \vec{\rho}_\mu \cdot \vec{\rho}^\mu} 
- {\frac{1}{4} \vec{R}_{\mu\nu} \cdot \vec{R}^{\mu\nu}} \nonumber \\
&& 
{+ \frac{1}{2} \left( \partial_\mu \vec{\pi} \cdot \partial^\mu \vec{\pi} - m_\pi^2 \vec{\pi}^2 \right)}. \label{lagrangianm}
\end{eqnarray}

\noindent
${\text{where}}\:{W_{\mu\nu}} = 
{\partial_\mu \omega_\nu - \partial_\nu \omega_\mu}$ {and} ${\vec{R}_{\mu\nu}} = {\partial_\mu\vec{\rho}_\nu-\partial_\nu\vec{\rho}_\mu.}$ {Here $\sigma$ and $\omega$ are isoscalar mesons, and $\vec{\pi}$ and $\vec{\rho}$ are isovector meson.}

{The lagrangian of interaction $\mathcal{L}_{\text{int}}$ writes}
\begin{eqnarray}
{\mathcal{L}_{\text{int}}} &=& 
{\sum_B \bar{\psi}_B \Bigg[ 
g_{\sigma B}(\sigma)\sigma - g_{\omega B} \gamma_\mu \omega^\mu 
+ \frac{f_{\omega B}}{2\mathcal{M}} \sigma_{\mu\nu} \partial^\nu \omega^\mu} \nonumber \\
&& {
- g_{\rho B} \gamma_\mu \, \vec{\rho}^\mu \cdot \vec{I}_B 
+ \frac{f_{\rho B}}{2\mathcal{M}} \sigma_{\mu\nu} \partial^\nu \vec{\rho}^\mu \cdot \vec{I}_B} \nonumber \\
&& {
- \frac{f_{\pi B}}{m_\pi} \gamma_5 \gamma_\mu \, \partial^\mu \vec{\pi} \cdot \vec{I}_B 
\Bigg] \psi_B.} \label{lagrangianint}
\end{eqnarray}
{Here the common scale mass $\mathcal{M}$ is taken to be free nucleon mass, the commutation operator for the $\gamma$ matrix writes $\sigma_{\mu\nu}=i\frac{[\gamma_\mu,\gamma_\nu]}{2}$, $\vec{I}_B$ denotes the isospin matrix for baryon $B$, which is set to $\vec{I}_B=0$ for iso-singlet baryons. Note that $f_{\omega B}$ and $f_{\rho B}$ are isoscalar-tensor and isovector-tensor coupling constants, respectively. the terms involving $g_{\sigma B}(\sigma)$, $g_{\omega B}$, $g_{\rho B}$, and $f_{\pi B}$ correspond to $\sigma-$, $\omega-$, $\rho-$, and $\pi-B$ couplings, respectively. Also note that $g_{\omega B}$, $g_{\rho B}$, $f_{\omega B}$, $f_{\rho B}$, and $f_{\pi B}$ are constants. In this formulation, the meson field values are replaced by the constant mean-field values, i.e. $\bar{\sigma}$, $\bar{\omega}$, and $\bar{\rho}$ (the $\rho^{0}$ field).}

\begin{table}[ht]
\centering
\begin{tabular}{cccccc}
\toprule
{$B$} & {$N$} & {$\Lambda$} & {$\Sigma$} & {$\Xi$} \\
\midrule
{$g_{\sigma B}/\sqrt{4\pi}$} & {1.94} & {2.15} & {1.67} & {1.50} \\
{$g_{\omega B}/\sqrt{4\pi}$} & {2.39} & {2.82} & {2.82} & {2.09} \\
{$f_{\omega B}/\sqrt{4\pi}$} & {$-0.545$} & {$-3.39$} & {$-0.261$} & {$-4.40$} \\
{$g_{\rho B}/\sqrt{4\pi}$} & {0.596} & {0} & {1.19} & {0.596} \\
{$f_{\rho B}/\sqrt{4\pi}$} & {3.39} & {0} & {2.94} & {$-0.446$} \\
{$f_{\pi B}/\sqrt{4\pi}$} & {0.268} & {0} & 0.190 & {$-0.0772$} \\
\bottomrule
\end{tabular}
\caption{{Coupling constants used in this model.}} 
\label{coupling}
\end{table}

{The coupling constants used in this model is shown by Table \ref{coupling}, while the prediction of the model at saturation is presented in Table \ref{prediction}.}

{ 
\begin{table}[ht]
\centering
\begin{tabular}{ccccc}
\toprule
\multicolumn{1}{c}{$w_0$} & \multicolumn{1}{c}{$n_0$} & \multicolumn{1}{c}{$S_0$} & \multicolumn{1}{c}{$K_0$} & \multicolumn{1}{c}{$L$} \\
\multicolumn{1}{c}{(MeV)} & \multicolumn{1}{c}{(fm$^{-3}$)} & \multicolumn{1}{c}{(MeV)} & \multicolumn{1}{c}{(MeV)} & \multicolumn{1}{c}{(MeV)} \\
\midrule
${-16.1}$ & ${0.155}$ & ${33.6}$ & ${274}$ & ${77.1}$ \\
\bottomrule
\end{tabular}
\caption{Predictions of this model at saturation density, including saturation energy $w_0$, saturation density $n_0$ itself, symmetry energy $S_0$, incompressibility $K_0$, and slope of the symmetry energy $L$.} \label{prediction}
\end{table}}

{In this model \cite{Miy}, at the inner crust, a fraction of neutrons become unbound and "drip" out of the nuclei; while at the outer crust, neutrons remain bound within the nuclei. In the core, there exist hyperons, where the potential depth of the hyperons reads}
\begin{eqnarray}
U_{Y}^{(N)} &=& {-g_{\sigma Y}(\bar{\sigma}^{(N)}) \bar{\sigma}^{(N)} + g_{\omega Y} \bar{\omega}^{(N)}} \nonumber \\
&& {+ \frac{1}{2M_Y} \left[-g_{\sigma Y}(\bar{\sigma}^{(N)}) \bar{\sigma}^{(N)} + g_{\omega Y} \bar{\omega}^{(N)}\right]^2.}
\end{eqnarray}
{The hyperon species $\Lambda$, $\Sigma$, and $\Xi$ are collectively represented by $Y$. The quantities $\bar{\sigma}^{(N)}$ and $\bar{\omega}^{(N)}$ correspond to the mean-field values of the $\sigma$ and $\omega$ meson fields in symmetric nuclear matter, respectively. The details of this model—including the way to obtain the energy density—can be referred to Ref. \cite{Miy}. After one obtain the energy density $\varepsilon$, the pressure $p$ can be calculated as $p=n_B^2\frac{\partial}{\partial n_B}\left(\frac{\varepsilon}{n_B}\right),$ where $n_B$ denotes the total baryon number density. It is evident that the EOS model employed in this work incorporates physically realistic components}
\subsection{Rotational mass of anisotropic neutron stars}
\label{sec:3}
The metric of rotating NSs with HT approximation is written as \cite{Wen}

\begin{eqnarray}
    ds^2&=&-e^{2\varphi}[1+2(h_0+h_2P_2(\cos\theta))] dt^2\nonumber\\
    &&+\left[1+\frac{2}{r}\left(m_0+m_2P_2(\cos\theta)\right)\left(1-\frac{2m(r)}{r}\right)^{-1}\right]\left(1-\frac{2m(r)}{r}\right)^{-1}dr^2\nonumber\\
&&+\:r^2 \sin^2\theta[1+2(v_2-h_2)P_2(\cos\theta)](d\phi-\omega\:dt)^2\nonumber\\
&&+\:r^2 [1+2(v_2-h_2)P_2(\cos\theta)]d\theta^2,\label{metricrotstar}
\end{eqnarray}

Here, $\omega$ denotes the angular velocity of the local inertial frame, which is proportional to the star's angular velocity $\Omega$ relative to a distant observer. Both $\omega$ and $\Omega$ satisfy $\omega=\Omega-\bar{\omega}$, where $\bar{\omega}$ is the angular velocity of the star relative to the local inertial frame. Note that $h_0$, $h_2$, $m_0$, $m_2$, and $v_2$ are perturbation functions; $P_2(\cos\theta)$ is the second order of Legendre polynomial equation; $e^{2\varphi}$ is a metric function.

The anisotropic form of the effective EMT writes
\begin{eqnarray}
\tilde{T}^\mu_\nu=(\tilde{\varepsilon}+\tilde{q})u^\mu u_\nu+\tilde{q}\delta^\mu_\nu+\sigma\chi^\mu\chi_\nu.\label{Teffectiveaniso}
\end{eqnarray}
Here $u_\nu$ is the fluid 4-velocity, $\chi_\nu$ is a unit radial vector that satisfies $u^\mu \chi_\nu=0$, $q$ is the tangential pressure which satisfies $\sigma=p-q$, where $p$ is the radial pressure, $\tilde{\varepsilon}=\varepsilon+\left(\frac{\kappa\lambda}{4\kappa\lambda-1}\right)T$, $\tilde{p}=p-\left(\frac{\kappa\lambda}{4\kappa\lambda-1}\right)T$, and $\tilde{q}=q-\left(\frac{\kappa\lambda}{4\kappa\lambda-1}\right)T$. It is important to note that it is valid to assume that $T^r_\theta=0$ \cite{Sil,Pat,Mallick,Rizaldy,Bas}.

The function $e^{2\varphi}$ is constrained by
\begin{eqnarray}
\frac{d\varphi}{dr}=\frac{m+\frac{\kappa}{2}\tilde{p}}{r(r-2m)},\label{metricPhiRastall}
\end{eqnarray}
while the mass $m$ is constrained by
\begin{eqnarray}
\frac{dm}{dr}=\frac{\kappa}{2} r^2 \tilde{\varepsilon}.
\label{mass}
\end{eqnarray}

The anisotropic TOV equation within Rastall gravity reads
\begin{eqnarray}
\frac{dp}{dr}&=&\left[-\frac{\left(\varepsilon+p\right)\left(m+\frac{\kappa}{2}r^3 \tilde{p}\right)}{r(r-2m)}-\frac{2\sigma}{r}+\frac{2\kappa\lambda}{4\kappa\lambda-1}\frac{d\sigma}{dr}\right]\nonumber\\
&&\times\left[1-\frac{\kappa\lambda}{4\kappa\lambda-1}\left(3-\frac{d\varepsilon}{dp}\right)\right]^{-1}.
\label{aTOVR}
\end{eqnarray}
For rotating relativistic stars, the $\bar{\omega}(r)$, with an initial arbitrary central value $\bar{\omega}(r=0)=\omega_c$, can be determined by solving Eq.~(72) in Ref.~\cite{Pat}.

By calculating the Rastall field equation, for the ($tt$)-component of the $l=0$ order, one can obtain
\begin{eqnarray}
\frac{dm_0}{dr}&=&\frac{\kappa}{2}r^2 \frac{d\tilde{\varepsilon}}{dp}\left(1-\frac{d\sigma}{dp}\right)^{-1}\left(\varepsilon+p\right)\left(1-\frac{\sigma}{\varepsilon+p}\right)\mathcal{P}_0\nonumber\\
&&+\frac{1}{12}j^2r^4\left(\frac{d\bar{\omega}}{dr}\right)^2-\frac{1}{3}r^3\left(\frac{dj^2}{dr}\right)\bar{\omega}^2\left(1-\frac{\sigma}{\varepsilon+p}\right) \nonumber\\
&&\times\left(1-\frac{\kappa\lambda}{4\kappa\lambda-1}\right),
\label{m0}
\end{eqnarray}
where $j=e^{-\varphi}\left(1-\frac{2m}{r}\right)^{1/2}$. From the ($rr$)-component of the $l=0$ order, one can obtain
\begin{eqnarray}
\frac{d\mathcal{P}_0}{dr}&=&-\frac{m_0\left(1+\kappa r^2 \tilde{p}\right)}{\left(r-2m\right)^2}-\frac{\kappa r^2\left(\varepsilon+p\right)}{2\left(r-2m\right)}\left(1-\frac{\sigma}{\varepsilon+p}\right)\nonumber\\
&&\times\left(1-\frac{d\sigma}{dp}\right)^{-1}\beta\:\mathcal{P}_0+\frac{1}{12}\frac{r^4j^2}{r-2m}\left(\frac{d\bar{\omega}}{dr}\right)^2\nonumber\\
&&+\frac{1}{3}\left[\frac{d}{dr}\left(\frac{r^3j^2\bar{\omega}^2}{r-2m}\right)-\gamma\right],
\label{p_0}
\end{eqnarray} 
where $\beta=1+\frac{\kappa\lambda}{4\kappa\lambda-1}\left(1-\frac{d\varepsilon}{dp}\right)+\frac{2\kappa\lambda}{4\kappa\lambda-1}$ and $\gamma=\frac{r^3}{r-2m}\frac{\kappa\lambda}{4\kappa\lambda-1}\left(\frac{dj^2}{dr}\right)\bar{\omega}^2\left(1-\frac{\sigma}{\varepsilon+p}\right).$ Here $\mathcal{P}_0$ is the pressure perturbation factor. Remember that $m_0(r=0)=\mathcal{P}_0(r=0)=0$.

The angular momentum $J$ of NSs writes
\begin{eqnarray}
    J=\frac{R^3}{2}\left(\Omega-\bar{\omega}(R)\right).\label{angularmomentum}
\end{eqnarray}
From Eq. (\ref{angularmomentum}), the moment of inertia $I$ of NSs can be calculated, i.e. $I=\frac{J}{\Omega}$.

Mass correction of the stars is given by $\delta M=m_0(r=R)+\frac{J^2}{R^3}$,
where $R$ is radius of the stars. Finally, the rotational mass reads
\begin{eqnarray}
    M=M_{stat}+\delta M,
\end{eqnarray}
where $M_{stat}$ is the total mass of NS within static configuration which is obtained by solving Eq. (\ref{mass}) and Eq. (\ref{aTOVR}).

\subsection{Anisotropy}
\label{sec:4}
In this study, we use the anisotropic pressure model proposed by Horvat et al. \cite{Hor}. Mathematically, the model can be expressed as
\begin{eqnarray}
\sigma\equiv \zeta\left(\frac{2m(r)}{r}\right)p(r). \label{HIM}
\end{eqnarray}
Here $\zeta$ is a constant that represents the anisotropic stength. When the tangential pressure exceeds the radial pressure (i.e. $\zeta<0$), the anisotropy may be more efficient in supporting highly compact bodies against gravitational collapse than in the opposite case (i.e. $\zeta>0$) \cite{Hor}. 

\begin{figure*}
	\centering 
	\includegraphics[width=\textwidth, height=5cm]{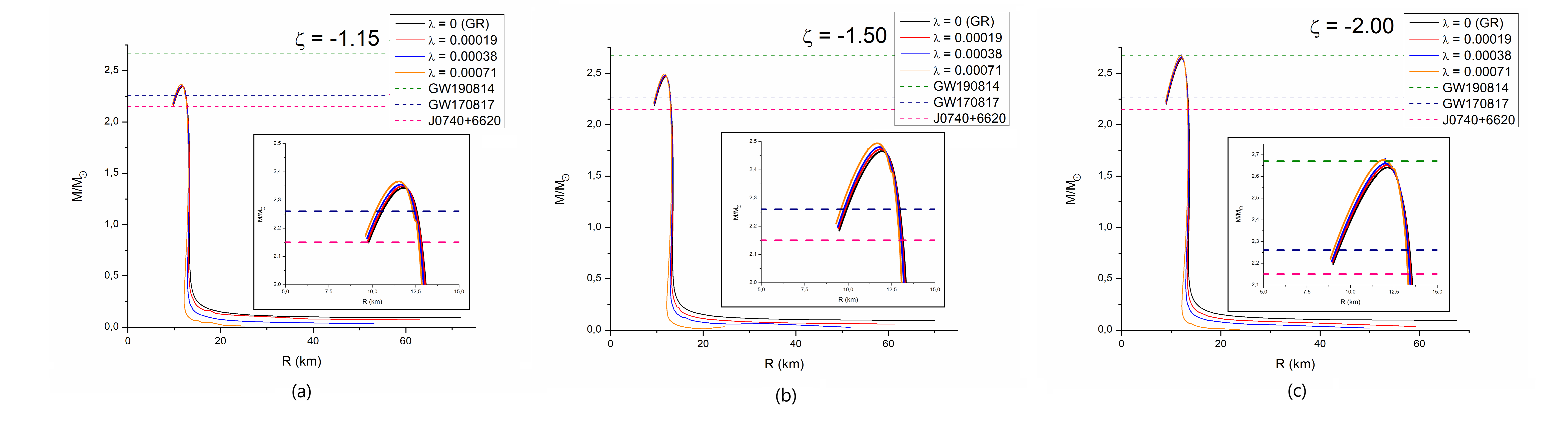}	
	\caption{Relation between radius and rotational mass of the anisotropic NSs with (a) $\zeta=-1.15$, (b) $\zeta=-1.50$, and (c) $\zeta=-2.00$. Those values agree with the values of $\zeta$ in Ref. \cite{Rah}. The angular velocity relative to the distant observers $\Omega =$ 1000 s$^{-1}$. For every $\zeta$ value, there are variations of the $\lambda$ values. The dashed lines are the upper limit of the mass constraints.} 
	\label{fig1}%
\end{figure*}
\begin{figure*}
	\centering 
	\includegraphics[width=\textwidth, height=5cm]{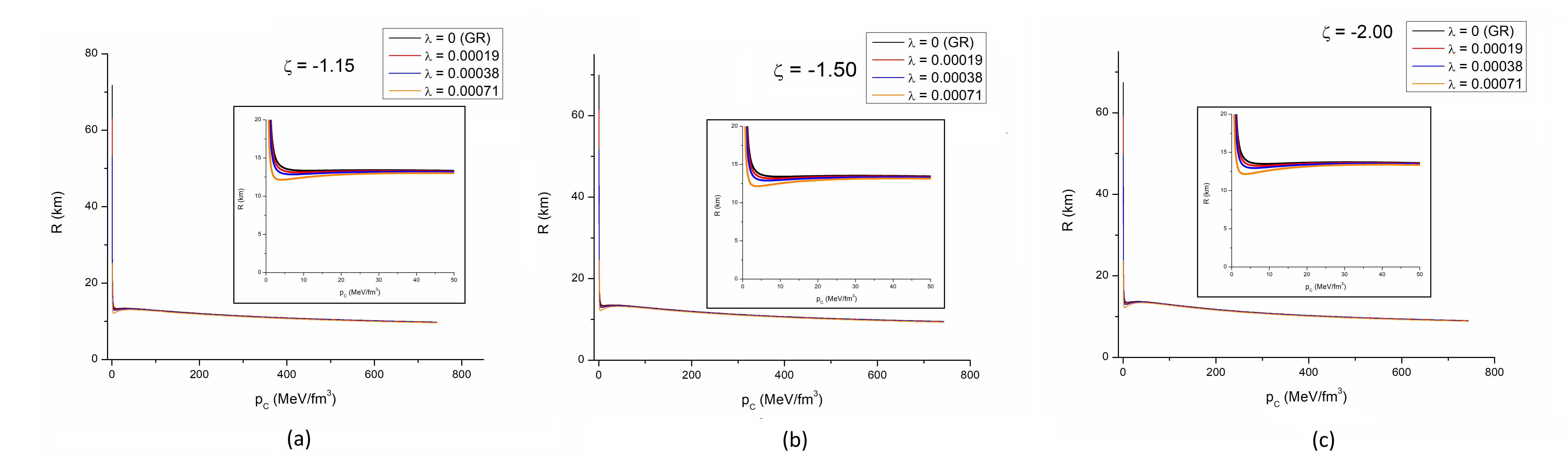}	
	\caption{Relation between pressure at the center of the NSs and rotational mass of the anisotropic NSs with (a) $\zeta=-1.15$, (b) $\zeta=-1.50$, and (c) $\zeta=-2.00$. The angular velocity relative to the distant observers $\Omega =$ 1000 s$^{-1}$. The variation of $\zeta$ values do not give obvious change to the NSs' radii; while from the variations of the $\zeta$ and $\lambda$, we can see that the increment of $\lambda$ tends to slightly reduce the radii of the NSs.} 
	\label{fig2}%
\end{figure*}
\begin{figure*}
	\centering 
	\includegraphics[width=\textwidth, height=5cm]{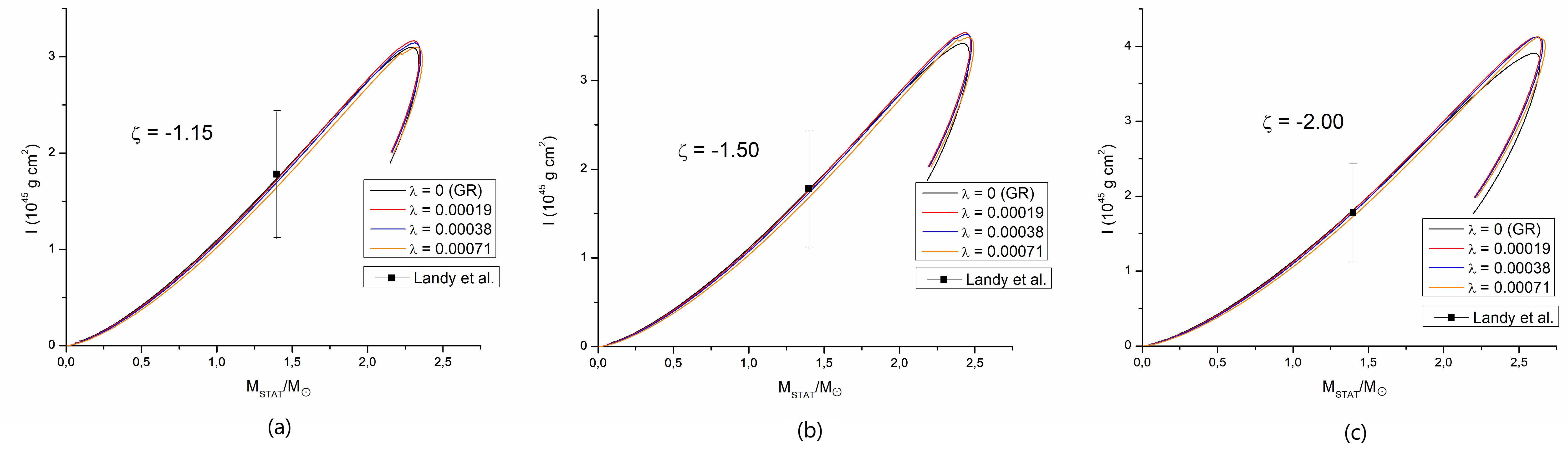}	
	\caption{Relation between rotational mass and moment of inertia of the anisotropic NSs with (a) $\zeta=-1.15$, (b) $\zeta=-1.50$, and (c) $\zeta=-2.00$. The angular velocity relative to the distant observers $\Omega =$ 1000 s$^{-1}$. The variations of anisotropic strength $\zeta$ impact on the increment of moment of inertia. The impact of $\lambda$ variations become more obvious as $\zeta$ increases. Here we use the constraint of moment of inertia from the study of Landry et al. \cite{Landryetal}} 
	\label{fig3}%
\end{figure*}
\begin{figure}
	\centering 
	\includegraphics[width=0.4\textwidth, angle=0]{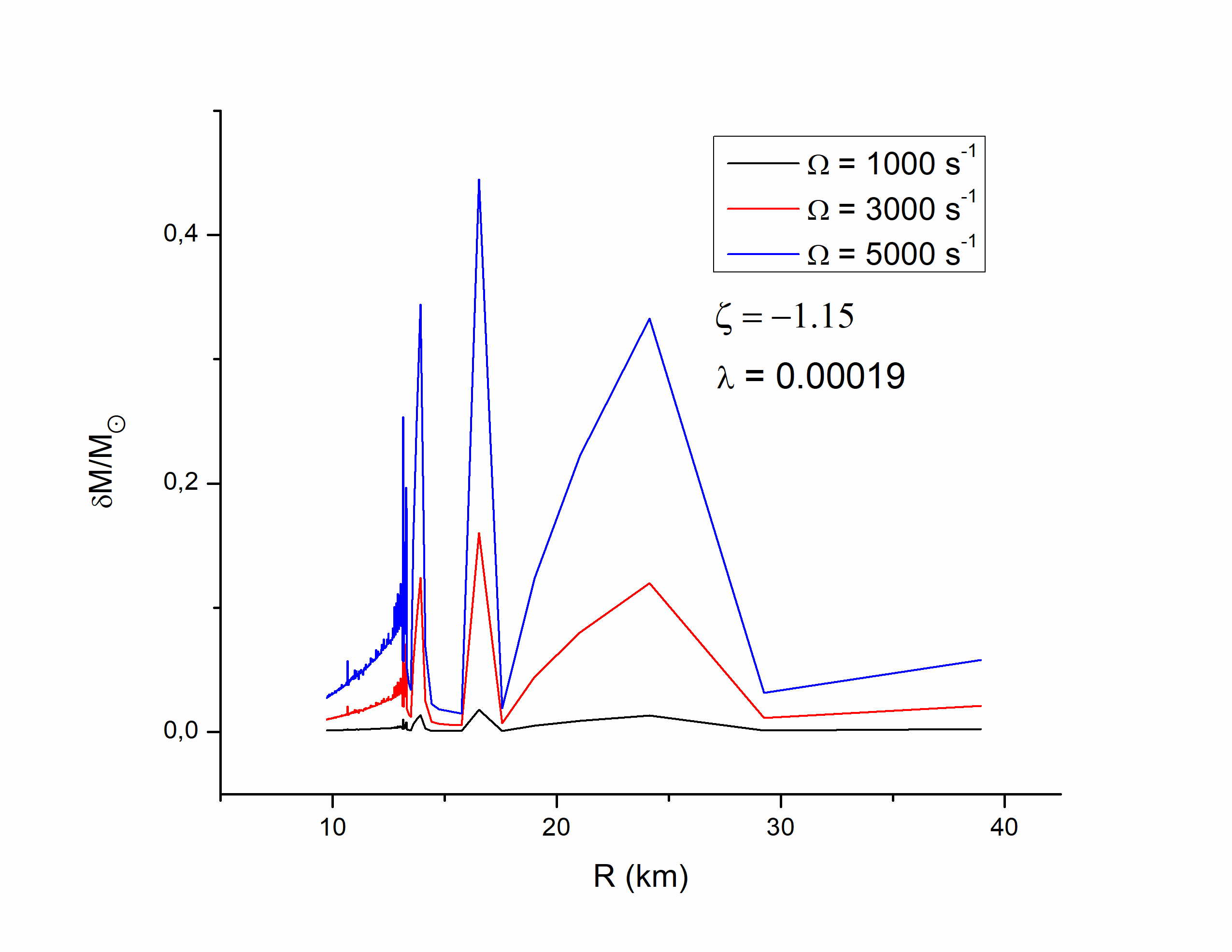}	
	\caption{Relation between radius and mass correction of anisotropic NSs when $\zeta$ and $\lambda$ are kept constant at $\zeta=-1.15$ and $\lambda=0.00019$. The angular velocity relative to the distant observers $\Omega$ is varied. The mass corrections significantly increase as $\Omega$ increases.} 
	\label{fig4}%
\end{figure}

\section{Numerical results and discussion}
\label{sec:5}
In the numerical calculations, we follow the algorithm of the one used by Ref. \cite{Ordaz}. We use Euler method to solve all differential equations simultaneously. For the EOS, we used the model from Ref. \cite{Miy}; with nuclei in the crust, and hyperons in the core. The main differential equations that have to be solved in this study are Eq. (\ref{mass}), Eq. (\ref{aTOVR}), Eq. (\ref{m0}), and Eq. (\ref{p_0}). The boundary conditions are $m(r=0)=0$, $m_0(r=0)=0$ and $\mathcal{P}_{0}(r=0)=0$; while $p(r=0)=p_c$ and $p(r=R)=0$, where $p_c$ is the pressure at the center of the NSs. The important entities are the values of $m_0(r=R)$ and $\mathcal{P}_0(r=R)$, which contribute to give the correction on NS's mass. They are obtained when calculations of differential equations stop, i.e. when $r=R$. It has to be noted that the differential equations that contain $m_0$ and $\mathcal{P}_0$ are coupled. We can see that $\mathcal{P}_0$ exists in the first term of the right-hand side of Eq. (\ref{m0}) which is used to obtain the function $m_0$; and we can find $m_0$ exist in the first term of the right-hand side of Eq. (\ref{p_0}) which is used to obtain the function $\mathcal{P}_0$. This fact demands us to simultaneously calculate Eq. (\ref{m0}) and Eq. (\ref{p_0}), even though we do not need the function $\mathcal{P}_0$ in the direct calculation of the rotational mass $M$ of NSs.

{The procedure to obtain $\bar{\omega}(r)$ and $\frac{d\bar{\omega}}{dr}$ is presented in Ref. \cite{Pat}. Note that Rastall parameter $\lambda$ does not affect the calculations of $\bar{\omega}(r)$ and $\frac{d\bar{\omega}}{dr}$.} Since the value of $\bar{\omega}_c$ as the initial value of $\bar{\omega}$ can be chosen arbitrarily, we choose 240 $s^{-1}$ in this work. It has to be highlighted that this value is much less than the value used in Ref. \cite{Wen}, i.e. $\bar{\omega}_c$ = 3000 s$^{-1}$. Choosing our value of $\bar{\omega}_c$ is also relatively wise since it is more suitable for HT formalism. We have to consider that $\bar{\omega}(r)$ increases as $r$ increases. Although we have a fact that according to Ref. \cite{Ber}, the HT formalism is accurate to better than 1 per cent even for the fastest millisecond pulsars, putting $\bar{\omega}_c$ = 3000 s$^{-1}$ in the calculation is potential to be awful, for the HT formalism is basically constructed for slowly rotating relativistic stars. So our value of $\bar{\omega}_c$ is safer than the one in Ref. \cite{Wen}. Comparing to Ref. \cite{Ordaz} and Ref. \cite{Pat} which chose $\bar{\omega}_c=0$, our choice of $\bar{\omega}_c$ is also more realistic for general rotating NS, for their choice tends to only agree the NSs with very slow rotation.

Fig. \ref{fig1} shows the relationship between radius and rotational mass of anisotropic neutron stars (NSs). The figure clearly demonstrates that Rastall gravity differs from GR ($\lambda = 0$). The Rastall parameter $\lambda$, can either increase or decrease the mass of the NS within specific ranges of $R$. In this study, we use three values of $\lambda$: $\lambda = 0.00019$, $\lambda = 0.00038$, and $\lambda = 0.00071$. {Table \ref{rastallparameter} displays the calculated values of $\kappa\lambda$ and $\Lambda$ for each value of $\lambda$. It is verified that the values of $\lambda$ used in this work do not violate any restrictions specified in Eqs. (\ref{RicciRastall}), (\ref{kappa}), (\ref{capitallambda}), (\ref{kappamoradpour}), and (\ref{lambdamoradpour}). Thus, all values of $\lambda$ employed in this work are allowed.}

\begin{table}
\centering
\begin{tabular}{ c c c} 
 \hline
 {$\lambda$} & {$\kappa\lambda$} & {$\Lambda$} \\ 
 \hline
{0.00019} & {0.08493} & {0.12862} \\ 
 {0.00038} & {0.08652} & {0.13230} 	 \\
 {0.00071} & {0.08928} & {0.13888} \\
 \hline
\end{tabular}
\caption{{The values of $\lambda$ and the corresponding values of $\kappa\lambda$ and $\Lambda$.}
}
\label{rastallparameter}
\end{table}

Each curve indicates that the $\lambda$ parameter affects the increase in the rotational mass of NSs within a specific regime, i.e., the region of the peaks of the curves. Since the mass values are high and the radii are small in this regime, it is characterized by high compactness. For each curve, these regimes are located in the range of $R = 9.78 - 12$ km for $\zeta = -1.15$, with a maximum compactness of 0.2263; in the range of $9.49 - 12.21$ km for $\zeta = -1.50$, with a maximum compactness of 0.2374; and in the range of $9 - 12.65$ km for $\zeta = -2.00$, with a maximum compactness of 0.2510. We refer to these regions as the compact regimes.

The $\lambda$ parameter also contributes to a decrease in rotational mass within another regime, specifically the region where the curves decline. These regimes are clearly discernible over the range from 15 km to the rightmost end of each curve, with minimum compactness values of 0.0005 for $\zeta = -1.15$, 0.0006 for $\zeta = -1.50$, and 0.0002 for $\zeta = -2.00$. We refer to these regions as the loose regimes.

It is interesting to compare the results shown in Fig. \ref{fig1} with those of da Silva et al. \cite{Das}, who also studied rotating NSs but under isotropic pressure. In their work, they used positive $\Lambda$ values, which correspond to negative $\lambda$ values. This causes the resulting pattern to be inverted; the compact regime in Ref. \cite{Das} corresponds to a decrease in mass, while their loose regime corresponds to an increase. It should be noted that the authors of Ref. \cite{Das} used $\Lambda$ instead of $\lambda$ in their calculations, and if we convert their $\Lambda$ values to $\lambda$, their absolute values of $\lambda$ are generally smaller than ours.

In the compact regime, the $\lambda$ parameter does not significantly affect the mass; while in the loose regime, the effect of the variations of $\lambda$ parameter on the mass becomes obvious. In term of obviousness of the $\lambda$ parameter's impact, the same pattern can be find in Ref. \cite{Das}.

In this study we use three values of anisotropic strength, i.e. $\zeta=-1.15$, $\zeta=-1.50$, and $\zeta=-2.00$. We can see that this anisotropy strength influences the mass increment more significantly than the $\lambda$ parameter. When $\zeta=-1.15$, the mass constraints of J0740+6620 and GW170817 can be achieved. NSs with $\zeta=-1.50$ also can achieve the mass constraint of J0740+6620 and GW170817. Unfortunately, these two values of anisotropic strength cannot achieve the constraint of GW190814. Interestingly, all constraints are achieved when $\zeta=-2.00$. We can see that the orange curve which is corresponds to $\lambda=0.00071$ in Fig. \ref{fig1}(a) can right coincide with the mass constraint of GW190814. It is such an interesting achievement that we can reach the mass value of an object which is still debatable whether it was a BH or a NS. If we revisit the results of da Silva et al. \cite{Das}, so we can get an insight that they also have achieved the mass constraint of GW190814 by varying the polytropic EOS, where they enlarged the EOS constant and lessened the polytropic index. However, it has to be noted that the EOS of Miyatsu et al. \cite{Miy} that we use is more realistic than the polytropic EOS. Moreover, according to the discussion in Ref. \cite{Das}, the EOS which can reach the mass constraint of GW190814 produces the mass-radius curves that are similar to the curves of quark stars in GR. Furthermore, the original maximum mass of NS generated by the EOS we use is 1.95 $M_\odot$, so adding rotation, anisotropy, and Rastall gravity to this simulation really impact the maximum mass of NS. These actualities lead to an argument that our model is more realistic for NS's model in Rastall gravity.

Fig. \ref{fig2} shows the relation between the pressure at the center $p_c$ and the radii of NSs. As we can see at the beginning of the curve, higher value of $\lambda$ parameter reduces the radius of NS. The anisotropic strength also has the same nature. We can see that as the anisotropic strength getting stronger, the radius slightly decrease. It is important to note that both $\lambda$ and $\sigma$ do not have significant impact on $p_c$ and radius relation. It is interesting to notice that this finding is different from the result in Ref. \cite{Pat}. Using the Bowers-Liang anisotropy model \cite{Bowersliang}, they find that at a regime, the radius increases as the anisotropy is getting stronger.

The moment of inertia is a significant physical quantity to be inspected, since it can be an effective way to ascertain the internal structure of the star, i.e. its EOS. Due to the connection between EOS and moment of inertia, it is possible to construct approximations for the moment of inertia so that this physical quantity can be estimated also in the study of static models, especially in the case of stiff EOS \cite{Das}. Moreover, a measurement of NS moment inertia is crucial because it has a universal relation with compactness \cite{Rah}. Figure \ref{fig3} illustrates the relationship between the static neutron star (NS) mass, $M_{stat}$, and the moment of inertia of NSs. It is evident that both the Rastall parameter $\lambda$ and the anisotropic strength $\zeta$ influence this relationship. Notably, the effect of $\lambda$ becomes more pronounced as $\zeta$ increases. Generally, the anisotropic strength magnifies the value of $I$; while Rastall's parameter decreases the moment of inertia in the regime before the peak, and increases the moment of inertia in the regime after the peak. We use the constraint of moment of inertia from the study of Landry et al. \cite{Landryetal}. The constraint is based on the lower limits on the maximum mass of NSs obtained through radio observations of heavy pulsars. We can see that all results match with the constraint's range. The optimal result is obtained when $\zeta = -2.00$, where the curves intersect the region surrounding the central value of the moment of inertia constraint.

We also need to analyze the impact of the angular velocity relative to the distant observers $\Omega$ on the mass correction due to rotation. The data that we plot in Fig. \ref{fig1}, Fig. \ref{fig2}, and Fig. \ref{fig3} are at $\Omega=1000$ s$^{-1}$. This condition we keep so that we can ensure that the basic property HT formalism as a slow rotation approach is still satisfied. In Fig. \ref{fig4}, we show the relation between radius of the stars and the mass correction $\delta M$. The plot is at $\lambda=0.00019$ and $\zeta=-1.15$. We can see that the increment of $\delta M$ is significant due to the increment of $\Omega$, and the increment of $\delta M$ is not linear to the increment of $\Omega$. This lead us to a fact that as $\Omega$ escalates, the rotational mass also significantly escalates. We can also take an insight from this condition that the increment of $\Omega$ impacts on the increment of the compactness of the stars.   




\section{Summary and conclusions}
This work extends the notion of rotational mass of anisotropic NS for the case of Rastall gravity. For the NS matter description, we use the EOS with nuclei in the crust, and hyperons in the core, proposed by Miyatsu et al. \cite{Miy}. While for the anisotropy profile, we use the model proposed by Horvat et al. \cite{Hor}. The parameter $\zeta$ in the anisotropy profile is interpreted as the anisotropic strength. The Rastall's parameter $\lambda$ interpretes the deviation of Rastall gravity from GR. When $\lambda=0$, the formulation is recovered to GR. Here we use three values of $\lambda$, i.e. $\lambda=0.00019$, $\lambda=0.00038$, and $\lambda=0.00071$. We also use three values of $\zeta$, i.e. $\zeta=-1.15$, $\zeta=-1.50$, and $\zeta=-2.00$. We also use three mass constraints, i.e. mass of J0740+6620, GW170817, and GW190814.

We have found that both $\lambda$ and $\zeta$ impact on the increment of the NS's rotational mass within the compact regimes, and also impact on the decrease of the NS's rotational mass within the loose regimes. The impact of $\lambda$ on the rotational mass of NSs is more obvious at the loose regimes. The factor that mostly escalates increment of the rotational mass is the anisotropic strength. All mass constraints are satisfied by the NS with $\zeta=-2.00$. On the other side, the higher value of $\lambda$ and the stronger value of anisotropic strength reduce the radius of NS. Moreover, the anisotropic strength magnifies the value of $I$; while Rastall's parameter decreases the moment of inertia in the regime before the peak, and increases the moment of inertia in the regime after the peak. All numerical results match with the constraint range of moment of inertia which is based on  the lower limits on the maximum mass of NSs obtained through radio observations of heavy pulsar. In term of the angular velocity of the stars relative to the distant observers $\Omega$, the mass correction $\delta M$ significantly increases when $\Omega$ increases, and the increment of the $\delta M$ is not linear to the increment of $\Omega$.

\section*{Acknowledgements}
We thank Anna Campoy Ordaz for making her code publicly available, which allowed MLP to modify it for use in this work. MLP gratefully acknowledges the Indonesia Endowment Fund for Education (LPDP) for financial support. FPZ would like to thank to the Ministry of Higher Education, Research, and Technology of the Republic of Indonesia (Kemendiktisaintek) through the LPPM ITB for partial financial support. MLP and HLP would also like to thank the members of the Theoretical Physics Groups at Institut Teknologi Bandung for their hospitality. MFARS is supported by the Second Century Fund (C2F), Chulalongkorn University, Thailand.

\section*{Declaration of generative AI and AI-assisted technologies in the writing process}

During the preparation of this work, M. Lawrence Pattersons used ChatGPT (OpenAI) in order to improve the language and readability of the manuscript. After using this tool, he reviewed and edited the content as needed and take full responsibility for the content of the publication.





\begin{thebibliography}{00}

\bibitem{Opp}
J.R. Oppenheimer, G.M. Volkoff, Phys. Rev. 55 (1939) 374.
\bibitem{Har1}
J.B. Hartle, Astrophys. J. 150 (1967) 1005.
\bibitem{Har2}
J.B. Hartle, K.S. Thorne, Astrophys. J. 153 (1968) 807.
\bibitem{Wen} D.H. Wen, W. Chen, L.G. Liu,
Chin. Phys. Lett. 22 (2005) 1604.
\bibitem{Weber}
F. Weber, N.K. Glendenning, Astrophys. J. 390 (1992) 541-549.
\bibitem{weber2}
F. Weber, N.K. Glendenning, Phys. Lett. B 265 (1-2) (1991) 1-5.
\bibitem{Ordaz}
A.C. Ordaz, Rotating Hyperonic Neutron Stars. Master's thesis, Universitat de Barcelona, 2019.
\bibitem{Urb}
M. Urbanec, J.C. Miller, Z. Stuchlík, Mon. Not. R. Astron. Soc. 433 (1013) 1903$-$1909.
\bibitem{Gup}
A. Gupta, S. Iyer, A.R. Prasanna, Class. Quantum Grav. 13 (1996) 2675, arXiv:gr-qc/9603055.
\bibitem{Pin}
E.O. Pinho, C.C. Barros Jr., Eur. Phys. J. C 83 (2023) 745.
\bibitem{Bez}
B. Bezděková, J. Bičák, Phys. Rev. D 108 (2023) 084043.
\bibitem{Moreno}
P.N. Moreno, F.J. Llanes-Estrada, E. Lope-Oter, Ann. Phys. 459 (2023) 169487, 	arXiv:2307.15366 [nucl-th].
\bibitem{Lopes}
L.L. Lopes, Astrophys. J. 966 (2024) 184.
\bibitem{Ber}
E. Berti, F. White, A. Maniopoulou, M. Bruni, Mon. Not. R. Astron. Soc. 358 (2005) 923-938.
\bibitem{Sil}
H.O. Silva, C.F.B. Macedo, E. Berti, L.C.B. Crispino, Class. Quantum Grav. 32 (2015) 145008.
\bibitem{Pat}
M.L. Pattersons, A. Sulaksono, Eur. Phys. J. C 81 (2021) 698.
\bibitem{Cook}
G.B. Cook, S.L. Shapiro, S.A. Teukolsky, Astrophys. J. 424 (1994) 823.
\bibitem{Mank}
V.S. Manko, E.W. Mielke, J.D. Sanabria-G\'{o}mez, Phys. Rev. D 61 (2000) 081501(R), arXiv:gr-qc/0001081.
\bibitem{Sul1}
A. Sulaksono, Int. J. Mod. Phys. E 24 (01) (2015) 1550007, arXiv:1412.7247 [nucl-th].
\bibitem{Her}
L. Herrera, N.O. Santos, Phys. Rep. 286 (2) (1997) 53-130.
\bibitem{Danarianto2019}M.D. Danarianto, A. Sulaksono, Phys. Rev. D 100 (2019) 064042.
\bibitem{Wojnar2016}A. Wojnar, H. Velten, Eur. Phys. J. C 76 (2016) 697.
\bibitem{Ovalle2017}J. Ovalle, Phys. Rev. D 95 (2017) 104019.
\bibitem{Afonso2018a}V. I. Afonso, G. J. Olmo, D. Rubiera-Garcia, Phys. Rev. D 97 (2018) 021503.
\bibitem{Afonso2018b}V. I. Afonso, G. J. Olmo, E. Orazi, D. Rubiera-Garcia, Eur. Phys. J. C 78 (2018) 866.
\bibitem{Bel}
P. Beltracchi, C. Posada, 	Phys. Rev. D 110 (2024) 024052, arXiv:2403.08250 [gr-qc].
\bibitem{Clifton2012}
  T. Clifton, P.G. Ferreira, A. Padilla, C. Skordis, 
  Phys. Rep. 513 (2012) 1$-$189, arXiv:1106.2476 [astro-ph.CO].
\bibitem{Ras}
P. Rastall, Phys. Rev. D 6 (1972) 3357.
\bibitem{Oli}
A.M. Oliveira, H.E.S. Velten, J.C. Fabris, L. Casarini, Phys. Rev. D 92 (2015) 044020.
\bibitem{Sak}
M.F.A.R. Sakti, A. Suroso, A. Sulaksono, F.P. Zen, Phys. Dark Univ. 35 (2022) 100974.
\bibitem{Vis}
M. Visser, Phys. Lett. B 782 (2018) 83-86.
\bibitem{Dar}
F. Darabi, H. Moradpour, I. Licata, Y. Heydarzade, C. Corda, Eur. Phys. J. C 78 (2018) 25.
\bibitem{Das}
F.M. da Silva, L.C.N. Santos, C.C. Barros Jr, Class. Quantum Grav. 38 (2021) 16.
\bibitem{Sak2}
M.F.A.R. Sakti, A. Suroso, F.P. Zen, Ann. Phys. 413 (2020) 168062, arXiv:1901.09163 [gr-qc].
\bibitem{Men}
L. Meng, D.J. Liu, Astrophys. Space Sci. 366 (2021) 105, arXiv:2111.03214 [gr-qc].
\bibitem{Tan}
T. Tangphati, A. Banerjee, S. Hansraj, A. Pradhan, Ann. Phys. 452 (2023) 169285.
\bibitem{Pre}
J.M.Z. Pretel, C.E. Mota, Gen. Rel. Grav. 56 (2024) 43, 	arXiv:2403.02440 [gr-qc].
\bibitem{Nas2}
G.G.L. Nashed, W. E. Hanafy, Eur. Phys. J. C 82 (2022) 679.
\bibitem{Nas}
G.G.L. Nashed, Universe 8 (2022) 10.
\bibitem{Tan2}
T. Tangphati, A. Banerjee, İ. Sakallı, A. Pradhan, Chin. J. Phys. 90 (2024) 422-433.
\bibitem{Sal}
I.G. Salako, D.R. Boko, G.F. Pomalegni, M.Z. Arouko, Theoret. Math. Phys. 208 (2021) 1299-1316.
\bibitem{Xi}
P. Xi, Q. Hu, G.N. Zhuang, X.Z. Li, Astrophys. Space Sci. 365 (2020) 163.
\bibitem{Pri}
H.L. Prihadi, M.F.A.R. Sakti, G. Hikmawan, F.P. Zen, Int. J. Mod. Phys. D 29 (03) (2020) 2050021, arXiv:1908.09629 [gr-qc].
\bibitem{Spa}
E. Spallucci, A. Smailagic, Int. J. Mod. Phys. D 27 (02) (2018) 1850003, 	arXiv:1709.05795 [gr-qc].
\bibitem{Hey}
Y. Heydarzade, H. Moradpour, F. Darabi, Can. J. Phys. 95 (12) (2017) 1253-1256, arXiv:1610.03881 [gr-qc].
\bibitem{Zou}
D.C. Zou, M. Zhang, C. Wu, R.H. Yue, Adv. High Energy Phys. 2020 (2020) 4065254.
\bibitem{Kumar}
R. Kumar, S.G. Ghosh, Eur. Phys. J. C 78 (2018) 750.
\bibitem{Heydar}
Y. Heydarzade, F. Darabi, Phys. Lett. B 771 (2017) 365-373.
\bibitem{Bha}
P. Bhar, Chin. J. Phys. 87 (2024) 782-796.
\bibitem{Malik}
A. Malik, A. Ashraf, F. Mofarreh, A. Ali, M. Shoaib, Int. J. Geom. Methods Mod. Phys. 20 (09) (2023) 2350145.
\bibitem{Hal}
S. Halder, S. Bhattacharya, S. Chakraborty, Mod. Phys Lett. A 34 (12) (2019) 1950095.
\bibitem{Mor}
H. Moradpour, N. Sadeghnezhad, S.H. Hendi, Can. J. Phys. 95 (12) (2017) 1257-1266, 	arXiv:1606.00846 [gr-qc].
\bibitem{Hey2}
Y. Heydarzade, M. Ranjbar, Eur. Phys. J. Plus 138 (2023) 703, arXiv:2307.04259 [gr-qc].
\bibitem{Bro}
K.A. Bronnikov, V.A.G. Barcellos, L.P. de Carvalho, J.C. Fabris, Eur. Phys. J. C 81 (2021) 395.
\bibitem{Aya}
A. Banerjee, T. Tangphati, S. Hansraj, A. Pradhan, Ann. Phys. 451 (2023) 169267.
\bibitem{Maj}
K. Majeed, G. Abbas, J. Phys. Commun. 6 (2022) 045005.
\bibitem{Cap}
M. Capone, V.F. Cardone, M.L. Ruggiero, J. Phys. Conf. Ser. 222 (2010) 012012.
\bibitem{Maj2}
A. Majeed, G. Abbas, M.R. Shahzad, New Astron. 102 (2023) 102039.
\bibitem{Abbott2020}
R. Abbott et al. (LIGO Scientific Collaboration and Virgo Collaboration), Astrophys. J. Lett. 896 (22) (2020) L44.
\bibitem{Cap2}
C.D. Capano, A.H. Nitz, Phys. Rev. D 102 (2020) 124070.
\bibitem{Tew}
I. Tews et al., Astrophys. J. Lett. 908 (1) (2021) L1, arXiv:2007.06057 [astro-ph.HE].
\bibitem{Cle}
S. Clesse, J. García-Bellido, Phys. Dark Univ. 38 (2022) 101111.
\bibitem{Lyu}
F. Lyu et al., Mon. Not. Royal Astron. Soc. 525 (2023) 4321-4328.
\bibitem{Hua}
K. Huang, J. Hu, Y. Zhang, H. Shen, Astrophys. J. 904 (2020) 39.
\bibitem{Dex}
V. Dexheimer, R.O. Gomes, T. Klähn, S. Han, M. Salinas, Phys. Rev. C 103 (2021) 025808.
\bibitem{Bis}
B Biswas, R. Nandi, P. Char, S. Bose, N. Stergioulas, Mon. Not. Royal Astron. Soc. 505 (2021) 1600-1606.
\bibitem{Hor}
D. Horvat, S. Ilijić, A. Marunović, Class. Quantum Grav. 28 (2) (2011) 025009, 	arXiv:1010.0878 [gr-qc].
\bibitem{Miy}
T. Miyatsu, S. Yamamuro, K. Nakazato, Astrophys. J. 777 (2013) 4.
\bibitem{Fonseca2021}
E. Fonseca et al., Astrophys. J. Lett. 915 (1) (2021) L12.
\bibitem{Abbottgw170817}
B.P. Abbott et al. (LIGO Scientific Collaboration and Virgo Collaboration), Phys. Rev. Lett. 119 (2017) 161101.
\bibitem{Rah} A. Rahmansyah, A. Sulaksono, A.B. Wahidin, A.M. Setiawan, Eur. Phys. J. C 80 (2020) 769.
\bibitem{Prasetyo}
I. Prasetyo, H. Maulana, H.S. Ramadhan, A. Sulaksono, Phys. Rev. D 104 (2021) 084029.
\bibitem{Moradpour2016}
H. Moradpour, I.G. Salako, Adv. High Energy Phys. 2016 (2016) 492796.
\bibitem{Mallick}
R. Mallick, S. Schramm, Phys. Rev. C 89 (2014) 045805, 	arXiv:1307.5185 [astro-ph.HE].
\bibitem{Rizaldy}
R. Rizaldy, A. Sulaksono, J. Phys. Conf. Ser. 1080 (2018) 012031
\bibitem{Bas}
L. Baskey, S. Das, F. Rahaman, Eur. Phys. J. C. 84 (2024) 92.
\bibitem{Bowersliang}
R.L. Bowers, E.P.T. Liang, Astrophys. J. 188 (1974) 657$-$665.
\bibitem{Landryetal}
P. Landry, R. Essick, K. Chatziioannou, Phys. Rev.D 101 (2020) 123007.



\end{thebibliography}


\end{document}